\documentclass[aps,prl,twocolumn]{revtex4-1}
\usepackage{epsfig}
\usepackage{graphicx}
\usepackage{amsmath}
\usepackage{bm}

\usepackage{xcolor}
\usepackage{hyperref}
\hypersetup{
     colorlinks   = true,
     linkcolor    = blue,
     citecolor    = blue
}

\begin{document}

\DeclareGraphicsExtensions{.eps,.EPS,.pdf,.PDF}

\title{Relaxation of the collective magnetization of a dense 3D array of interacting dipolar S=3 atoms
}
\author{Lucas Gabardos$^{1,2,\dagger}$, Bihui Zhu$^{3,4,\dagger}$, Steven Lepoutre $^{1,7}$, Ana Maria Rey $^{5,6}$, Bruno Laburthe-Tolra $^{2,1}$ and Laurent Vernac $^{1,2}$}

\affiliation{$^{1}$\,Universit\'e Paris 13, Laboratoire de Physique des Lasers, F-93430, Villetaneuse, France\\
$^{2}$\,CNRS, UMR 7538, LPL, F-93430, Villetaneuse, France\\
$^{3}$\,ITAMP, Harvard-Smithsonian Center for Astrophysics, Cambridge, MA 02138, USA\\
$^{4}$\,Department of Physics, Harvard University, Cambridge, MA 02138, USA\\
$^{5}$\,JILA, NIST and Department of Physics, University of Colorado, Boulder, USA\\
$^{6}$\,Center for Theory of Quantum Matter, University of Colorado, Boulder, CO 80309, USA\\
$^{7}$\,Université Paris-Saclay, CNRS, ENS Paris-Saclay, Laboratoire Aimé Cotton, 91405, Orsay, France $^{\dagger}$ These authors contributed equally.
}

\begin{abstract}

We report on measurements of the dynamics of the collective spin length (total magnetization) and spin populations  in an almost unit filled lattice system comprising about $10^4$ spin $S=3$ chromium atoms, under the effect of dipolar interactions. The observed  spin population dynamics is unaffected  by the use of a spin echo, and fully consistent with numerical simulations of the $S=3$ XXZ spin model. On the contrary, the observed spin length decays slower than in simulations, and surprisingly reaches a small but nonzero asymptotic value within the longest timescale.
Our findings show that spin coherences are sensitive probes to systematic effects affecting quantum many-body  behavior that cannot be diagnosed by merely measuring spin populations.
\end{abstract}
\date{\today}
\maketitle

Synthetic atom-based materials are emerging as unique quantum laboratories for the exploration of  collective behaviors in  interacting many-body  systems\cite{Gross2017}. In particular  both electric and magnetic dipolar gases featuring   long range spin-spin interactions are opening  great opportunities  for the exploration of quantum magnetism   in regimes inaccessible to gases interacting via purely contact interactions \cite{Bohn2017}.

While electric  dipolar interactions are fundamentally stronger and have led to important breakthroughs as demonstrated by recent experiments using  KRb molecules in 3D  lattices \cite{Yan2013} and Rydberg  atoms in  bulk gases \cite{Signoles2019} as well as in optical tweezers arrays \cite{Bernien2017,Browaeys2020,Browaeys2019,Schau2015,Schau2012,Keesling2019},   magnetic quantum dipoles offer complementary unique opportunities for quantum simulations. For example, they provide the  possibility to trap  low entropy and dense macroscopic  arrays of $S>1/2$ atoms   in close to unit filled 3D optical lattice potentials     where truly collective many-body behavior manifests itself. Under these conditions it is possible  to study  spin models with large spins \cite{Lahaye2009,depaz2013,depaz2016}  which cost exponentially more resources to classically simulate \cite{Hallgren2013} than conventional $S=1/2$ models of magnetism. These capabilities have  started to be explored   in experiments working both with bosonic chromium and fermionic erbium atoms in 3D lattices \cite{depaz2013,depaz2016,Lepoutre2019,Fersterer2019,Patscheider2020}, which  have observed already signatures of  rich   many-body dynamics  including quantum thermalization  and the buildup of many-body correlations. However,  so far all the  information has only been extracted from measurements of spin populations without direct access to quantum coherences, which contain key signatures of the underlying quantum dynamics \cite{Yan2013,Signoles2019}.

 \begin{figure}
\centering
\includegraphics[width= 0.42\textwidth]{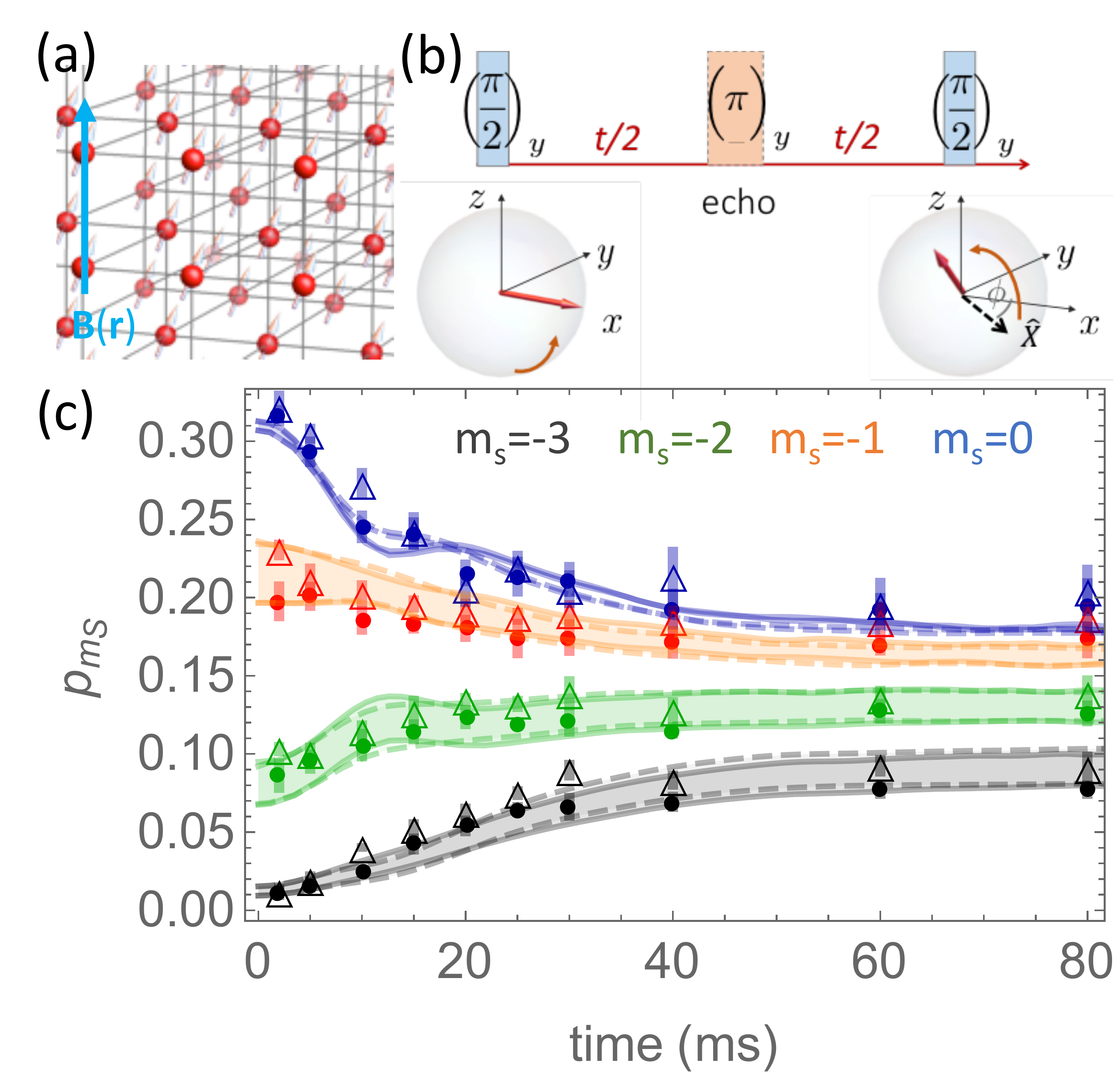}
\caption{\setlength{\baselineskip}{6pt} {\protect\scriptsize (a) The experimental system consists of a 3D array of dipolar Cr atoms, with a spatially varying magnetic field $B({\bf r})$. (b) Experimental RF sequence: Cr atoms are initialized in the $m_s=-3$ spin state and rotated by the first $\pi/2$ pulse to align with the $x$ direction; the 2nd $\pi/2$ pulse is used to measure the spin length. Due to magnetic noise the collective spin length makes an angle $\phi$ with respect to Ox when this second pulse is imparted. We impart or not an echo pulse using an additional $\pi$ pulse at half the evolution time $t/2$.  (c) Time-evolution of the fractional populations $p_{m_s}$ of 4 $m_s$ spin components, with and without spin echo (full circles and empty triangles respectively). Each data point corresponds to the average of 10 realizations. Error bars correspond to statistical uncertainties. The solid (dashed) lines show the numerical results obtained with GDTWA for spin dynamics with (without) spin echo, for a lattice with unit filling and $B_Q/h=-4$~Hz. From bottom to top: $m_s=-3$ (black),$m_s=-2$ (green), $m_s=-1$ (orange), $m_s=0$ (blue). The colored bands in the GDTWA calculations account for a $5\%$ experimental  error in the first pulse area.}}
\label{FigDyna}
\end{figure}

 Here we make a step forward and report time-resolved measurements   of the  spin coherence and also populations of a many-body strongly interacting    spin $S=3$ dipolar gas of $^{52}$Cr atoms in a deep 3D lattice. The spin coherence is extracted by measurements of the collective transverse magnetization of the gas, ${J_\bot}$,  via Ramsey spectroscopy. Since the longitudinal magnetization remains zero at all times, the measurement of the transverse magnetization can also be seen as a measurement of the total magnetization (or the collective spin length) of the ensemble.
 The system is initially prepared in a far-from-equilibrium spin  coherent state with maximal transverse magnetization   ${J_\bot}=SN$, which is let to evolve due to magnetic dipolar couplings.
Our experimental protocol includes a spin-echo pulse at the middle of the dynamics to reduce the effect of magnetic field inhomogeneities on the transverse magnetization dynamics.

In agreement with previous results \cite{Lepoutre2019,Fersterer2019,Patscheider2020}, we find that the  spin population dynamics is well captured by  a semiclassical method, referred to as the generalized discrete
truncated Wigner approximation (GDTWA), based on a discrete Monte Carlo sampling in phase space  \cite{Zhu_2019,Schachenmayer2015}. In addition, we find that spin dynamics is barely affected by the spin echo.  However,
we observe that the observed transverse magnetization not only decays at a slower rate than the one expected from a pure spin XXZ model  but also saturates at a non-zero value, behavior that is   inconsistent with  numerical expectations. We attribute the difference to  effects not included in the pure spin model such as tunneling  induced by lattice heating. We provide toy model simulations that support this claim. Our observations highlight the relevance of quantum coherence to characterize many-body phenomena.

Our experimental platform differs from previous studies on the transverse magnetization of ensembles of dipolar particles \cite{Yan2013,Signoles2019} in that it consists of a high density  \textit{ordered} array of \textit{S=3} spin  particles. It is obtained by loading a $^{52}$Cr BEC in a 3D optical lattice deep into the Mott regime. We obtain typically $N_S=10^4$ atoms close to unit filling (see \cite{Lepoutre2019} for details). Initially the sample is prepared in a spin coherent state, with all spins in the maximally stretched state $m_S=-3$ and aligned with the external magnetic field $\textbf{B}$. Spin dynamics is triggered by aligning all spins at $t=0$ along a direction orthogonal to $\textbf{B}$, by the use of a resonant RF pulse (See Fig. 1).  We measure the dynamical evolution of the seven spin populations $N_{m_S}$ in the basis set by the external magnetic field  through Stern-Gerlach separation. We  also probe the  collective spin length (total magnetization) $\mathbf{J}=\sum_{i=1}^{N_S}\mathbf{S}_i$ which has a norm   $\left|\left|\mathbf{J}\right|\right|= \sqrt{ \left<J_x\right>^2+\left<J_y\right>^2+\left<J_z\right>^2}$
ranging from $0$ to $3N_S$. The maximal value is reached for the maximally polarized state, e.g. the initial spin state. In the following we will use normalized quantities:  $\mathbf{j}=\mathbf{J}/N_S$ and $\ell=\left|\left|\mathbf{j}\right|\right|$.

To measure $\ell$ we  use a Ramsey protocol, in which a second $\pi/2$ rotation is imparted just before population measurements (See Fig. 1). In the rotating frame (turning around $z$ at the RF frequency), RF pulses ensure rotation of the spins around an axis called $y$. After the first $\pi/2$ pulse, $\mathbf{j}_{t=0} \propto \hat{x}$. During the spin dynamics, fluctuations of the external magnetic field make $\mathbf{j}$ rotate in the $xy$ plane. We denote $\phi(t)$ the angle between $\mathbf{j}$ and $\hat{x}$ and use it to   define a new basis $XYz$ where $\mathbf{j} \parallel \hat{X}$ i.e. $\ell=<\hat{j}_X>$. Since the second $\pi/2$ pulse again rotates spins around $y$, the normalized magnetization measured by the Stern and Gerlach protocol, denoted as $M_z$, corresponds to a measurement of $\hat{j}_{x}=\cos\left(\phi\right)\hat{j}_X-\sin\left(\phi\right)\hat{j}_Y$ after a Ramsey sequence. Since $\phi(t)$ is different  trial to trial, this random phase generates a net  dephasing which is useful to extract the net spin length.

If one can neglect  tunneling, the prepared ensemble of $N_S$ coupled spins, which are pinned at the individual sites  of a 3D lattice, evolve under a pure spin model. In the presence of an external magnetic $B$ field strong enough to generate Zeeman splittings
larger than nearest-neighbor dipolar interactions, the dynamics is described by the following XXZ spin model \cite{depaz2013}:
\begin{equation}
\hat H_{\rm dd}=\sum_{i> j}^{N_S} V_{ij} \left[ \hat S_i^z \hat S_j^z -\frac{1}{2} \left( \hat S_i^x \hat S_j^x + \hat S_i^y \hat S_j^y \right) \right]
\label{secular}
\end{equation}
$V_{i,j}= \frac{\mu_0 (g \mu_B)^2}{4 \pi} \left( \frac{1-3 \cos ^2 \theta _{i,j}}{r_{i,j}^3}\right)$, with $\mu_0$ the magnetic permeability of vacuum, $g \simeq 2$ the Land\'e factor, and $\mu_B$ the Bohr magneton. The sum runs over all pairs of particles ($i$,$j$). $r_{i,j}$ is the distance between atoms, $\theta_{i,j}$ the angle between their inter-atomic axis and the external magnetic field assumed to be along the $z$ axis, and  ${\bf{\hat S}}_i=\{\hat S_i^x,\hat S_i^y,\hat S_i^z \}$ are spin-3 angular momentum operators, associated with atom $i$ \cite{som}. For an ensemble of dipolar spins, the normalized spin length $\ell$ decreases as a result of interactions, which at short time follows the form:
\begin{equation}
\ell(t)=3-\frac{81t^2V_{\rm{eff}}^2}{16\hbar ^2},
\label{pert}
\end{equation}
where $V_{\rm{eff}}/h=\sqrt{1/N_S\times\sum_{i\neq j}^{N_S}V_{ij}^2}/h\simeq6$ Hz for a unit-filled lattice in our experiment. This leads to a typical timescale $\tau_{\rm dd}\simeq20$ ms for $\ell$ to reach 0. This is a pure quantum effect since  a mean-field ansatz predicts no decay \cite{Kawaguchi2007,LepoutrePRA2018,Lepoutre2019}.
We note that similar dipolar induced magnetization  decay and evidence of the  build up  of multiple-spin coherences  has been reported in NMR systems where  nevertheless the system starts in a highly mixed state \cite{NMR}.

\begin{figure}
\centering
\includegraphics[width= 8.8 cm]{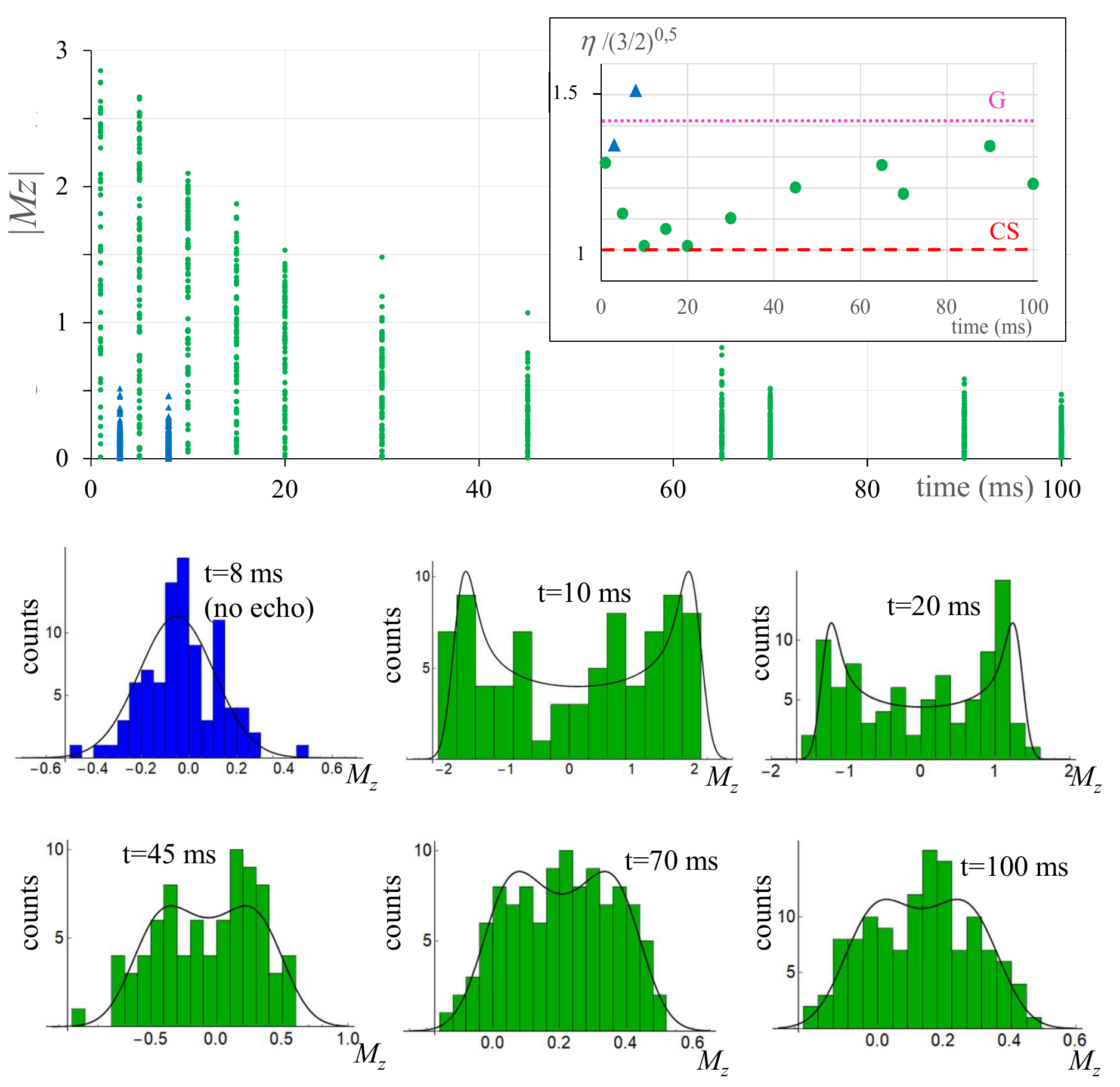}
\caption{\setlength{\baselineskip}{6pt} {\protect\scriptsize Measurement of the distribution of the normalized magnetization $M_z$ following a Ramsey sequence. Each time corresponds to 60 to 100 realizations. Top: Absolute values of $M_z$ are plotted for different spin dynamics durations; blue triangles and green circles correspond to experiments without and with spin echo respectively. Inset: The parameter $\eta$ (see text) is evaluated from the corresponding $M_z$ distributions. The horizontal lines show the expected value for respectively the PD of a classical spin (dashed, red), and the Gaussian PD (dotted, purple). Bottom: Histograms of the $M_z$ distributions are shown together with the PD used to fit them (solid line); see text.}}
\label{RawData}
\end{figure}

In addition to $\hat H_{\rm dd}$, atoms experience a tensor light shift $\hat { H}_{Q}=\sum_{i} B_{\rm Q} ({\hat S} _i^{z})^2$. For a non interacting gas this leads to a periodic evolution of $\ell$, with a time scale $\tau_q=\frac{h}{4|B_{\rm Q}|}\approx50$ ms to reach zero for typical $|B_{\rm Q}|/h \approx 5$ Hz in our experiment. This one-body term has to be taken into account in simulations. At short time, it leads to a replacement of  $V_{\rm{eff}}\rightarrow V_{\rm eff}\sqrt{1+40B_Q^2/27V_{\rm eff}^2}$ in Eq. \ref{pert}, thus making the decay of $\ell$ even  faster.

Furthermore, magnetic field inhomogeneities described by gradients for the Larmor frequency, $\omega_{\rm L,i}=g\mu_B/\hbar \left(B_0+\vec{b}.\vec{r_i}\right)$, lead to another term in the Hamiltonian, $ {\hat H}_{B}=\sum_{i}\hbar \omega_{\rm L,i} ({\hat S} _i^{z})$, which generates dephasing and leads to a damping of $\ell$. The damping timescale is $\tau_b=\frac{h}{2g\mu_B b R}\simeq3$ ms with $R\simeq 5$ $\mu$m the typical size of the sample, which is shorter than $\tau_{\rm dd}$. In order to compensate for this dephasing, we implement a spin-echo technique, in which spins are rotated by $\pi$ in the middle of the dynamics (see Fig.~\ref{FigDyna}).

One question that naturally arises is whether the spin echo changes as well the evolution of the populations $p_{m_S}$ of the  different spin components. As shown in Fig.~\ref{FigDyna}, the observed spin dynamics is roughly identical with and without the echo, which is confirmed by   GDTWA numerical simulations using the experimental gradient of (10.5$\pm1$) Gauss.m$^{-1}$. This behavior is consistent with a short time perturbative analysis, which predicts that magnetic field gradients only enter at quartic order in the population dynamics, i.e.  $p_{m_S}(t)-p_{m_S}(0)\propto \frac{t^4}{N_S}[15\sum_{i=1}^{N_S}(\sum_{j\neq i }^{N_S}   V_{ij} (\omega_{Li}- \omega_{Lj}))^2-27\sum_{i,j\neq i}^{N_S}(\omega_{Li}-\omega_{Lj})^2V_{ij}^2] $   while dipolar effects enter at second order $\propto t^2 V_{\rm eff}^2$~
\cite{som}.

Our raw experimental results for the measurement of the spin length are shown in Fig. \ref{RawData} (top). Without a spin echo, a fast damping of the magnetization is observed, in a timescale  consistent with $\tau_b$. There is here a striking difference with our previous measurements in a bulk BEC \cite{LepoutrePRA2018}, where a gap due to spin-dependent interactions prevents the reduction of magnetization. When a spin echo is applied, the raw data show that $\ell$ decays with a significantly longer timescale, compatible with $\tau_{dd}$. Note nevertheless that given that $\hat{H}_{dd}$ does not commute with $\hat{H}_{B}$ the utility of a spin-echo  to protect the decay of $\ell$, is parameter and geometry dependent \cite{Solaro2016}.

To obtain a quantitative estimate of $\ell$ as a function of time, we have investigated the probability distributions (PD) associated with the data. Figure \ref{RawData} shows that a mostly Gaussian PD is obtained for experiments without echo. On the contrary, data with spin-echo only show a Gaussian-like shape at long times. At short time PDs of a totally different kind are obtained, with a maximum of the probability for large values of $|M_z|$. To account for this observation, we introduce the probability distribution of a classical spin (CS) of norm $\ell$. Such PD is obtained by differentiating the projection $M_z=\ell\cos(\phi)$, thus obtaining the number $dN$ of realization of $M_z$:
\begin{equation}
\frac{dN}{dM_z}_{CS}=\frac{1}{\pi l}\frac{1}{\sqrt{1-\frac{M_z^2}{l^2}}}
\label{PDclass}
\end{equation}

\begin{figure}
\centering
\includegraphics[width=0.45\textwidth]{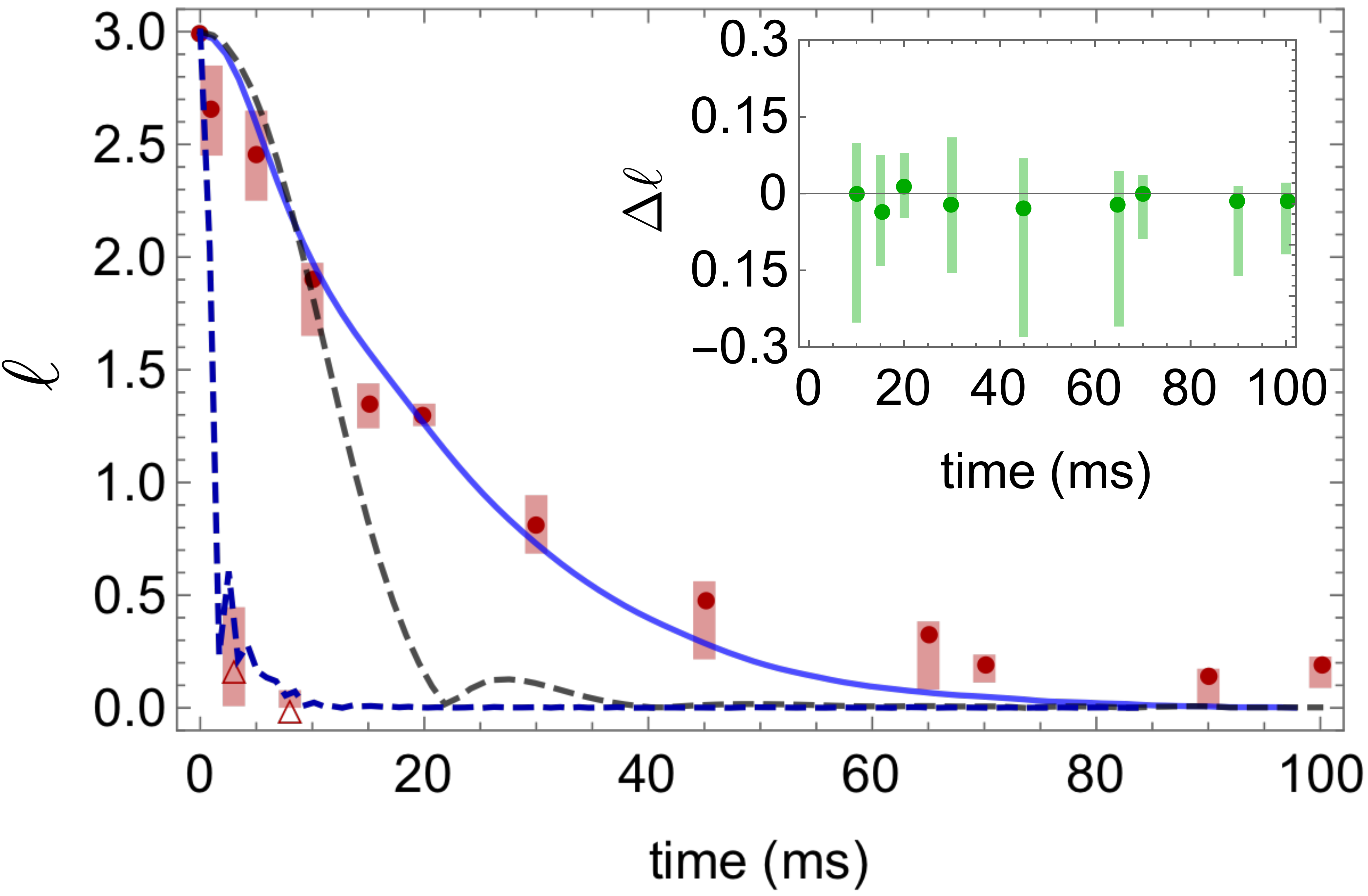}
\caption{\setlength{\baselineskip}{6pt} {\protect\scriptsize  Values of the spin length $\ell$ derived by fitting the distributions of the magnetization $M_z$ after the Ramsey sequence. Error bars represent the 68 $\%$ confidence interval, and are detailed in \cite{som}. Filled circles and empty triangles are measurements with and without the spin echo pulse, respectively. The black dashed line shows the numerical results with the spin echo pulse applied, obtained with GDTWA  for the same lattice configuration as in Fig.~\ref{FigDyna}. The blue dashed line shows the dynamics without the spin echo, obtained from the gaussian TWA simulations \cite{som}. The blue solid line corresponds to GDTWA simulations effectively accounting for atomic motion in the lattice, with $B_Q/h=-2$Hz.  Inset: difference between the two experimental determinations of $\ell$, comparing the results of the fit of the experimental probability distributions to Eq. (\ref{ell2}). Error bars correspond to the quadratic average of the standard deviations associated with either methods. }}
\label{FigLS}
\end{figure}

In order to characterize the observed PDs, we evaluate from the data the square-root of the kurtosis $\eta=\frac{\sqrt{M_4}}{M_2}$ with $M_n=\int PD(x)x^ndx$: $\eta=\sqrt{\frac{3}{2}}$ for the PD of eq.(\ref{PDclass}), and $\eta=\sqrt{3}$ for a Gaussian PD, $\frac{dN}{dM_z}_{G}=\frac{1}{\sqrt{\pi}\sigma}\exp\left(-\frac{M_z^2}{\sigma^2}\right)$. The experimental values of $\eta$ are shown in Fig.~\ref{RawData} . Data without echo show a good agreement with a Gaussian PD. For data with a spin-echo, the value of $\eta$ is in good agreement with the classical value for $t_f\simeq5-30$ ms, and it gradually approaches a gaussian value for $t>60$ ms. This first qualitative analysis shows trends for the measured PDs. In order to get numerical values of $\ell$ we have used a convolution of the two PDs described above to fit the data, as shown in Fig.~\ref{RawData}; this method assumes that a total dephasing has occurred, which requires $t \geq 10$ ms in our experiment (for $t=1,5$ ms we use another analysis, see \cite{som}).

The corresponding results of $\ell$ for each time $t$ are shown in Fig.~ \ref{FigLS}. As expected, without applying the echo pulse $\ell$ decays rapidly; the actual damping rate depends on the system size and lattice geometry. On the other hand, the measured data of $\ell$ after applying the echo pulse reveal an exponential damping of the collective spin towards a small but not zero value, $\ell(t)\simeq(3-\ell_0)\exp(-t/\tau_e)+\ell_0$, with $\ell_0=0.15$ and $\tau_e=22$ ms. This contrasts with the glassy dynamics observed in $e.g.$ \cite{Signoles2019}.

In order to model the dynamics of $\ell$ in the absence of the echo pulse, it is crucial to  appropriately account for the actual sample geometry in experiment and to capture the  effects of inhomogeneities. For this purpose, we implement a gaussian TWA approach in our numerical calculation (blue dashed line in Fig.~\ref{FigLS})~\cite{som}\footnote{Note: in this work, the gaussian TWA approach is only used for finding the time evolution of $\ell$ without the echo pulse [Fig.~\ref{FigLS} blue dashed line]; the other results are obtained with GDTWA. Also see \cite{som} for more details.}, which allows for efficiently simulating systems with $N_s\sim10^4$, much larger than the size previously investigated~\cite{Lepoutre2019}.  When the echo pulse is applied, we first use GDTWA simulations using the same parameters as those used  in Fig.~\ref{FigDyna}. We explicitly insert a $\pi$ rotation around the $y$ axis at half of the evolution time \cite{som}. While the GDTWA captures the populations dynamics at all times (see Fig.~\ref{FigDyna}), it is only able to reproduce the spin length measurements at $t \leq 10$ ms (black dashed line in Fig.~\ref{FigLS}). Interestingly, the  spin length dynamically evolves for $t>30$ ms whereas the population  dynamics and pure spin model numerical simulations  have then essentially reached a plateau. This indicates that measuring the collective spin length constitutes a more sensitive probe than simply monitoring spin dynamics.

While tunneling in the lowest band (where atoms  are initially loaded) is too slow to explain the discrepancy between the spin-length data and the GDTWA simulations, one possibility could be that phase noise in the lattice  could promote particles to higher bands, where tunneling is non-negligible. This type of heating processes was for example also reported with KRb molecules \cite{Chotia2012}. To model this possible scenario we performed  numerical simulations assuming frozen atoms but relaxing the requirement  to   be pinned  in the regular grid imposed by the lattice potential while keeping the same average density \cite{som}. This emulates the idea  that during a tunneling process on average an atom can be in between two adjacent  lattice sites. The calculated dynamics of spin population resulting from this toy model is consistent with the experimental measurements \cite{som}. The result for $\ell$ is shown with a  solid line in Fig.~\ref{FigLS}: the agreement with the experimental data is much better than the one obtained with GDTWA (black dashed line); however, our toy model predicts a zero relaxation value of the spin length within the experimental time range investigated, in contrast with the experimental observations.

In order to confirm our measurements of $\ell$, we performed a noise analysis of the components of the collective spin. Whether we apply the final $\pi/2$ pulse (Ramsey experiment, labelled \textit{R}) or not (experiment labelled \textit{noR}), we measure $\hat{j}_{x}$, or $\hat{j}_{z}$. Taking into account the randomness of $\phi$ \cite{Lucke2014} and a technical noise on the measurements, one obtains the following expressions for the variance of $M_z$ when averaging over many realizations:

\begin{eqnarray}
\rm{Var}(M_z)_{\rm{R}}&=&\frac{\ell^2}{2}+\frac{\rm{Var}(\hat{j}_X)+\rm{Var}(\hat{j}_Y)}{2}+\sigma_{\rm{exp}}^2\equiv\left<j_{x}^2\right>_{\rm{exp}}\nonumber\\
\rm{Var}(M_z)_{\rm{noR}}&=&\frac{3}{2N_s}+\sigma_{\rm{exp}}^2\equiv\left<j_{z}^2\right>_{\rm{exp}}
\label{Varjphijz}
\end{eqnarray}

To derive Eq.(\ref{Varjphijz}) we add a technical noise (with an associated standard deviation $\sigma_{\rm{exp}}$) to the theoretical expectations. Since  our main source of technical noise corresponds to  an insufficient signal to noise ratio in the absorption images, measurements of $\hat{j}_z$ and $\hat{j}_{x}$ are affected by the same technical noise. Since $\left[\hat H_{\rm dd},\hat{J}_z\right]=0$, the theoretically expected $\left<\hat{j}_{z}^2\right>$ remains equal to its value at $t=0$, i.e. to the standard quantum noise (SQN) $\frac{3}{2N_s}$.

Therefore measurements of standard deviations without Ramsey pulse ($\sqrt{\left<j_{z}^2\right>_{\rm{exp}}}$) brings a benchmark of the technical noise: our data show that we obtain about $3$ times the SQN for $t=0$, and that the ratio to SQN increases as a function of time \cite{som}. We therefore can consider that $\left<j_{z}^2\right>_{\rm{exp}}\simeq\sigma_{\rm{exp}}^2$. The technical noise can be compared with the values predicted by our simulations for the quantum noise of $\hat{j}_{x}$~\cite{som}: we get $\sigma_{\rm{exp,min}}^2>4\times \frac{\rm{Var}(\hat{j}_X)+\rm{Var}(\hat{j}_Y)}{2}_{\rm{max}}\simeq\frac{32}{N_s}$.  We therefore assume that $\left<j_{x}^2\right>_{\rm{exp}}\simeq\frac{\ell^2}{2}+\sigma_{\rm{exp}}^2$, so that:
\begin{equation}
\left<j_{x}^2\right>_{\rm{exp}}\simeq\frac{\ell^2}{2}+\left<j_{z}^2\right>_{\rm{exp}}
\label{ell2}
\end{equation}
Therefore, a signature of a non-zero $\ell$ is that  $\left<j_{x}^2\right>_{\rm{exp}}$ is larger than the technical noise. This provides a simple measurement of $\ell$ complementary to the method described above, provided that a complete dephasing happens (for $t\geq 10$ ms in our case). The two methods are in good agreement, as shown in the inset of Fig.~\ref{FigLS} .

{\it In conclusion}, our experiment demonstrates the remarkably different effects of a spin echo  on the dynamics of a strongly interacting quantum system for spin population and spin magnetization. Notably, our measurements show that the decay of the transverse magnetization in our experiment is slower than expected, and approaches a small but finite value at long times. This surprising observation indicates that our experiment cannot be fully described by a spin model of frozen particles, a finding that could not be previously deduced from the measurements on population dynamics. This illustrates how measurements of spin coherences provide valuable information on quantum many-body systems that are crucial to benchmarking experiments as quantum simulators. Our work also paves a way towards further investigations using spin coherences to probe quantum many-body phenomena in $S>1/2$ dipolar systems.

\begin{acknowledgments}
 We  thank Benoit Darqui\'e, Tommaso Roscilde and Johannes Schachenmayer for useful discussions, and Thomas Bilitewski and Itamar Kimchi
 for reviewing the manuscript. {\bf Funding:}   The Villetaneuse group acknowledges financial support from CNRS, Universit\'e Sorbonne Paris Cit\'e (USPC), Conseil R\'egional d’Ile-de-France under Sirteq Agency, the Indo-French Centre for the Promotion of Advanced Research - CEFIPRA under the LORIC5404-1 contract, Agence Nationale de la Recherche (project ANR-18-CE47-0004), and QuantERA ERA-NET (MAQS project). A.M.R is supported by the AFOSR grant FA9550-18-1-0319, by  the DARPA DRINQs grant, the ARO single investigator award W911NF-19-1-0210,  the NSF PHY1820885, NSF JILA-PFC PHY-1734006 grants, and by NIST. B.Z. is supported by the NSF through a grant to ITAMP.
\end{acknowledgments}

\bibliography{biblioLS}

\section{Supplemental Material}

\subsection{Extensive data: Standard deviations}

\begin{figure}
\centering
\includegraphics[width= 9 cm]{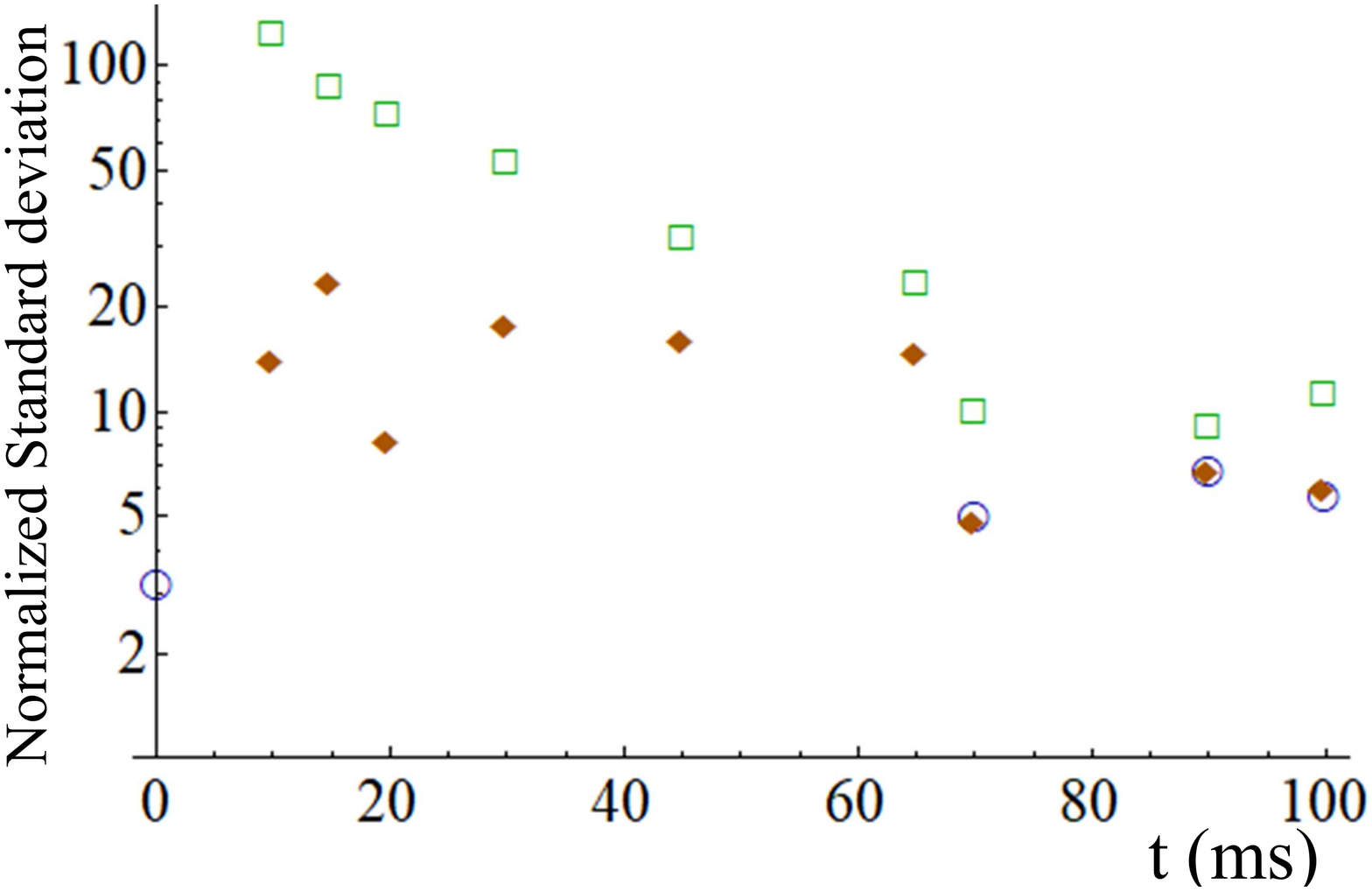}
\caption{\setlength{\baselineskip}{6pt} {\protect\scriptsize Normalized standard deviations to the standard quantum limit: experimental values for $j_x$ (open squares) and $j_z$ (open circles); and standard deviation $\sigma_G$ of the Gaussian derived from the convolution fit of the $M_z$ histograms (full diamonds).
 }}
\label{FigSD}
\end{figure}

We show in Fig \ref{FigSD} standard deviations normalized to the standard quantum limit (SQL) $\sqrt{\frac{3}{2N}}$. The experimental values $(\left<j_{x}^2\right>_{\rm{exp}}^{1/2}$ normalized to SQL are plotted for all times where total dephasing has occurred (see below). We plot as well our measured values of $(\left<j_{z}^2\right>_{\rm{exp}}^{1/2}$ normalized to SQL. During the data taking, we could minimize the technical noise by optimizing the intensity of the probe laser used for absorption imaging; this optimized intensity was used for data at long times, $t=70,90,100$ ms. For these three values of $t$ we measured both $j_x$ and $j_z$; we measured as well $j_z$ for $t=0$ ms, which corresponds to our minimal noise value, but still larger than SQL. We plot as well the normalized standard deviation obtained by our convolution fit analysis (which is the parameter $\sigma_G$ below); the good agreement between this standard deviation and the experimental one of $j_z$ shows the consistency of our analysis. Data corresponding to $10\leq t\leq65$ ms are affected by a larger technical noise, hence the discontinuity appearing between $t=65$ and $t=70$ ms.

\subsection{Estimate of $\ell$ at short time and extended data}
As explained in the main article, measurements of $\ell$ are based on two methods, which both require that a full dephasing has occurred, $i.e.$ that the angle $\phi$ between $\mathbf{j}$ and $\hat{x}$ (see main article) uniformly spans $\left[-\pi,+\pi\right]$. For data at short time dephasing has not taken place yet, which leads to non symmetric histograms (for $t=1$ ms and $t=5$ ms).

\begin{figure}
\centering
\includegraphics[width= 9 cm]{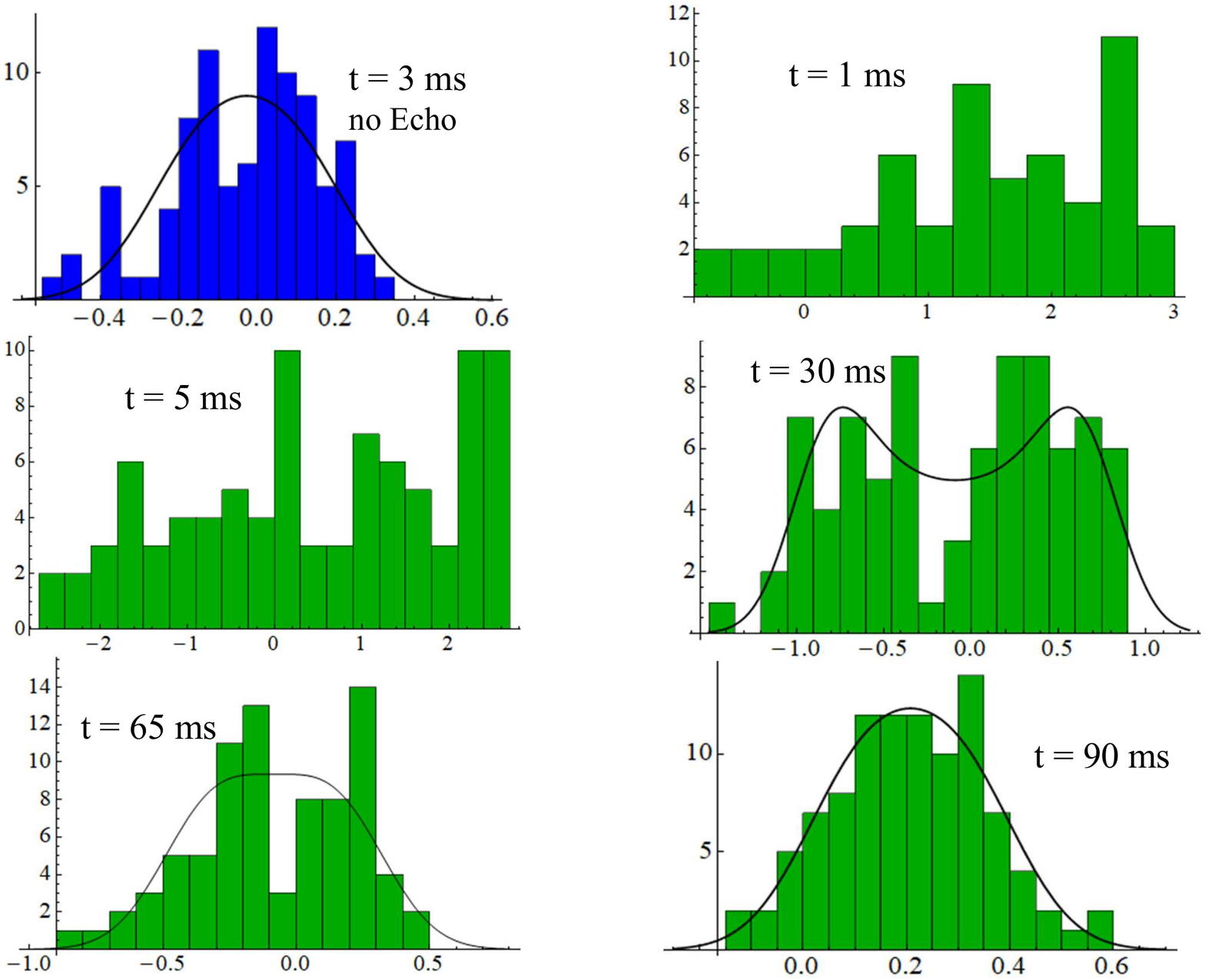}
\caption{\setlength{\baselineskip}{6pt} {\protect\scriptsize  Additional histograms of the $M_z$ distributions not shown in the main article, with the PD used to fit them when they are relevant.}}
\label{FigHistoSM}
\end{figure}

We show in Fig \ref{FigHistoSM} some  additional  histograms of the $M_z$ distributions not shown in the main article, with the PD used to fit them when it is relevant: as explained above for $t=1$ ms and $t=5$ ms non total dephasing prevents to fit the data.

\begin{figure}
\centering
\includegraphics[width= 8 cm]{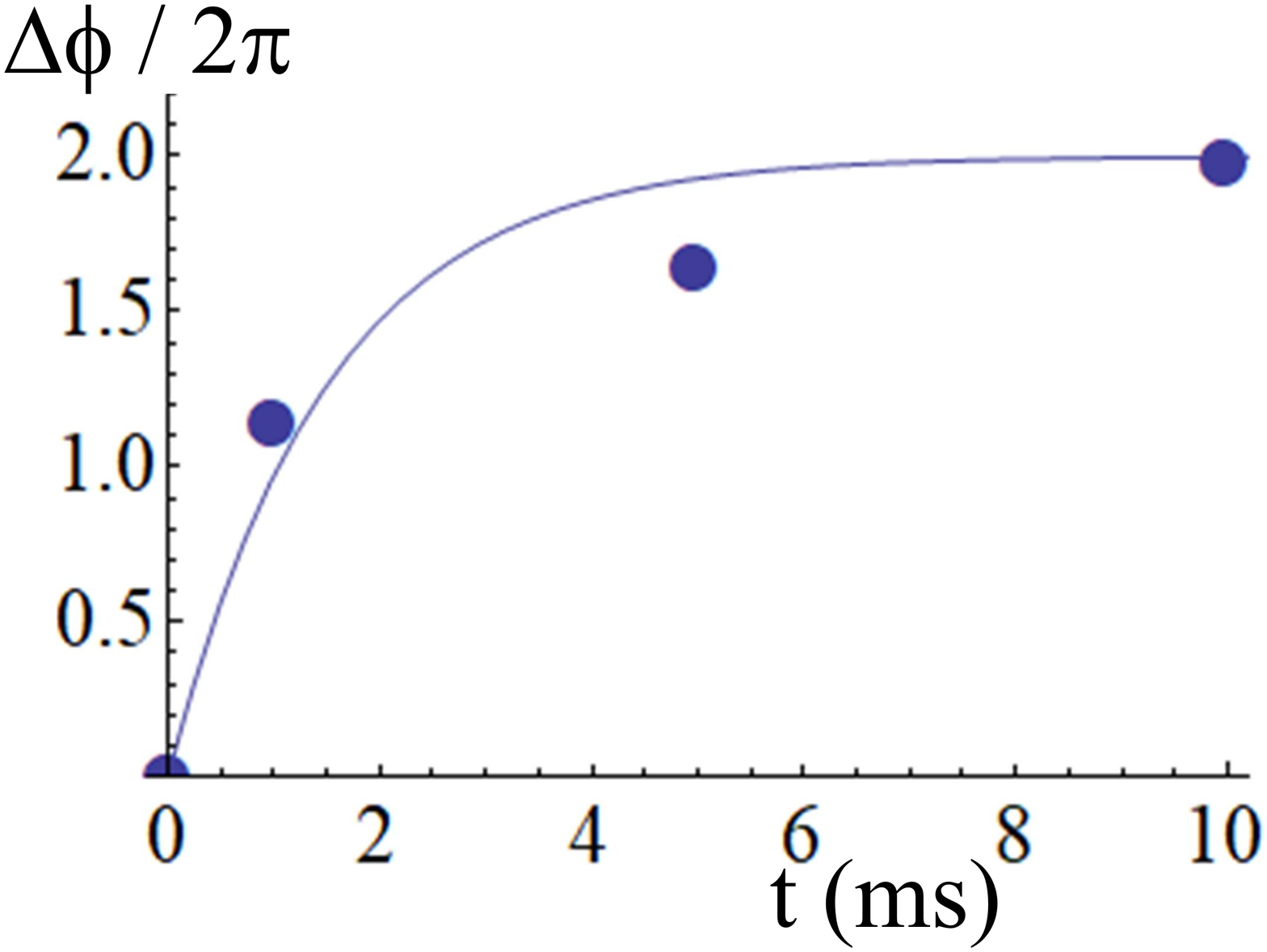}
\caption{\setlength{\baselineskip}{6pt} {\protect\scriptsize Estimate of the growth of the dephasing $\Delta_{\phi}$ with time in the experiment. The full line is an exponential to guide the eye.
 }}
\label{FigDphi}
\end{figure}

 We explain here how we evaluate the value of $\ell$ for these two short time values. We assume a classical spin, with a random dephasing angle $\phi$ ranging in the interval $-\frac{\Delta_{\phi}}{2}<\phi<\frac{\Delta_{\phi}}{2}$. Assuming a uniform distribution within this interval, we generate ensembles of values of $\ell \cos(\phi)$ with $N_{\rm{exp}}$ terms, where $N_{\rm{exp}}$ is the number of values of $M_z$ in the experimental data. We then look for the value of $\Delta_{\phi}$ and $\ell$ which lead to the same mean value and standard deviation as the ones of the experimental data: this provides our estimate of the central value of $\ell$, $\ell_c$. To roughly estimate the error bars, we assume $2\ell_c-M_{z\rm{max}}\leq\ell\leq M_{z\rm{max}}$, with $M_{z\rm{max}}$ the maximal experimental value of $M_z$: this very likely overestimates the error bar, as for all values of $t\geq10$ ms the value of $M_{z\rm{max}}$ is beyond error bars on $\ell$. We obtain in addition an estimate of the way the dephasing $\Delta_{\phi}$ grows in the experiment (see Fig \ref{FigDphi}).

\subsection{Error bars on $\ell$}
We now describe the way error bars are evaluated for data at $t \geq 10$ ms, when full dephasing has occurred. As explained in the main text, we estimate $\ell$ from two methods.

We start with the method using fit of the experimental Probability Distributions (PDs). We search for the PD which has the same second and fourth moments as the experimental data; these are denoted respectively by  $M_{2,\rm{exp}}$ and $M_{4,\rm{exp}}$. This distribution results from a convolution of a Gaussian (characterized by its standard deviation $\sigma_G$) and a Classical Spin distribution (characterized by the spin length $\ell_0$). The central value of $\ell$ is given by $\ell_0$. We use 50 $\%$ sub-sampling of the experimental data to estimate uncertainties on $M_{2,\rm{exp}}$ and $M_{4,\rm{exp}}$. We then vary the values of $M_2$ and $M_4$ in the intervals provided by the sub-sampling analysis, and obtain from convoluted PDs our estimate of $\ell_{min}$ and $\ell_{max}$, respectively the minimal and maximal value of $\ell$. We assume that the interval $[\ell_{min};\ell_{max}]$ defines the two-standard-deviations confidence interval for $\ell$.

We also performed the following analysis to better ascertain the non-zero value of $\ell$ at long times.
For that we generate numerically large number of samples with a Gaussian statistics, with the same number of values of $M_z$ (noted $N_{exp}$), the same mean value and standard deviation $\sqrt{M_2}$ as the ones of the experimental sample. We then obtain a list of values of the parameter $\eta = \frac{\sqrt{M_4}}{M_2}$ (see main text) with a mean value close to $\sqrt{3}$, and a standard deviation $\sigma_{eta}$. It gives us an estimate of the probability that the experimental sample is compatible with a pure Gaussian PD.
For example at $t=70$ ms, we obtain $\eta_{exp}=1.45$ for the experimental $M_z$ data, and $\sigma_{eta}=0.12$ ($N_{exp}=106$). Using this analysis, the probability that the experimental PD is compatible with a pure Gaussian is given by $\int_{-\infty}^{-(\sqrt{3}-\eta_{exp})/\sigma_{\eta}}\frac{1}{\sqrt(\pi)}\exp(-x^2)dx=0.01$. We find this probability to be $0.27$ for $t=90$ ms, and $0.03$ for $t=100$ ms. Since a Gaussian distribution is necessarily obtained for $\ell=0$ (assuming a negligible quantum noise), our analysis can exclude $\ell =0$ with a very high confidence for $t=70,100$ ms (while $\ell =0$ cannot be fully excluded at 90 ms).

We now turn to the determination of $\ell$ by the second method described in the article, see eq.(5). For $t=70,90,100$ ms we simply apply the formula with measured values of $\left<j_{x}^2\right>_{\rm{exp}}$ and $\left<j_{z}^2\right>_{\rm{exp}}$ (see Fig \ref{FigSD}). For $5\leq t\leq65$, we take for $\left<j_{z}^2\right>_{\rm{exp}}$ the variance of the data with no echo at $t=8$ ms: the corresponding probability distribution being very close to a Gaussian, we assume that its variance corresponds to the technical noise. As for error bars for this second method, uncertainties on $\left<j_{x}^2\right>_{\rm{exp}}$ and $\left<j_{z}^2\right>_{\rm{exp}}$ are evaluated through sub-sampling, which yields standard deviation on these quantities to be about $5\%$.

\subsection{Theoretical model}

We describe the experimental system with the Hamiltonian
\begin{eqnarray}
\hat H=\sum_{\bf r} B({\bf r}) \hat S^z_{\bf r}+B_Q\sum_{\bf r}(\hat S^z_{\bf r})^2\nonumber \\
+\frac{V({\bf r},{\bf r}')}{2}\sum_{{\bf r}\neq {\bf r}'}[\hat  S^z_{\bf r}\hat S^z_{{\bf r}'}-\frac{1}{2}(\hat  S^x_{\bf r}\hat S^x_{{\bf r}'}+\hat  S^y_{\bf r}\hat S^y_{{\bf r}'})]
\label{subeq:H}
\end{eqnarray}
where ${\bf r}$ denotes the coordinates of the lattice sites, and  $\hat S^{x,y,z}_{\bf r}$ are spin-3 angular momentum operators acting on atoms located at ${\bf r}$ \footnote{$\hbar$ is dropped in these definitions.}. For atoms frozen at discrete lattice sites, ${\bf r}=i_x d_x\hat x+i_y d_y\hat y+i_z d_z\hat z$ for the ${\bf i}$th atom. Here $d_{x,y,z}$ are lattice constants along the different directions.  $V({\bf r},{\bf r}')$ denotes the dipolar interactions between two atoms located at ${\bf r}$ and ${\bf r}'$, and its explicit form is  given in the main text between arbitrary ${\bf i}$ and ${\bf j}$ atoms. $B(\bf r)={\bf b}\cdot{\bf r}$ accounts for the inhomogeneous magnetic fields in the experiment.

To obtain the spin dynamics accounting for actual experimental conditions, we utilized the GDTWA approach introduced in Ref.~\cite{Zhu_2019} to numerically solve the time evolution under Eq.~(\ref{subeq:H}). For the case where an echo pulse is applied, we first numerically evolve the system under $\hat H$ for $t/2$, then implement a $\pi$ rotation around the $y$ axis, and evolve the system under $\hat H$ for another $t/2$ afterwards, to find the resulting dynamics at time $t$. Namely, the whole time evolution operator is $\hat{\mathcal{U}}_{\rm echo}(t,0)=e^{-i\hat Ht/2\hbar}e^{-i\pi\hat S^y}e^{-i\hat Ht/2\hbar}$, with $\hat S^y=\sum_{\bf r}\hat S_{\bf r}^y$. In Fig.~1 of the main text, we first simulated  the population dynamics of the different spin components $p_{m_s}(t)$ for a unit-filled lattice with $(L_x,L_y,L_z)$ number of sites along $(x,y,z)$ directions, using experimental values of $d_{x,y,z}$ and ${\bf b}$, for both the cases with and without a spin echo pulse. Since $d_y\approx 2 d_z\approx 2 d_x $, we used $(L_x,L_y,L_z)=(14,7,14)$ in our calculation. The resulting dynamics shows a convergence for such system sizes.  Since $B_Q$ is not exactly known in experiment, it was chosen  as a fitting parameter bounded by experimental estimation of $|B_Q|/h\leq 6$Hz. It was held fixed when comparing the echo and no echo cases. We find a good agreement between the numerics and experimental data, with small differences between the two cases.

As mentioned in the main text, a useful way to understand this behavior is to examine the  short-time quantum dynamics. Utilizing the series expansion of the evolution operator $\hat{ \mathcal{U}}=e^{-i\hat H t}=\sum_{k}\frac{(-i)^kt^k}{k!}\hat H^k$, we can express $p_{ms}(t)$ in different powers of $t$. We find that even without the echo pulse, the leading contribution to the population change  $\delta(p_{ms})=p_{ms}(t)-p_{ms}(0)$ comes from interactions. To provide an explicit description of the dynamics, here we focus on the change for $m_s=0$, for which the leading contribution is \footnote{$B_Q$  also affects $\delta(p_{ms})$ at this order, as shown in  Ref.~\cite{Lepoutre2019}.}
\begin{eqnarray}
\delta(p_{0})&\propto-\frac{135t^2}{128}V_{\rm eff}^2
\end{eqnarray}
and is not affected by the inhomogeneity ${\bf b}$. The effect of ${\bf b}$ shows up in the dynamics at later time as
\begin{eqnarray}
\delta(p_{0})=\Gamma \frac{5t^4}{256N}
\end{eqnarray}
with $\Gamma=15\sum_{{\bf r}}[\sum_{{\bf r}'\neq {\bf r} }  V({\bf r},{\bf r}') ({\bf b}\cdot({\bf r}-{\bf r}')]^2-27\sum_{{\bf r},{\bf r}'\neq {\bf r}}({\bf b}\cdot({\bf r}-{\bf r}'))^2V({\bf r},{\bf r}')^2$.
Similar effects also apply to other $m_s$ states and we refer the interested readers to Ref.~\cite{Lepoutre2019} for relevant details. In contrast, the dynamics of $\ell$ is much more sensitive to the echo pulse and the lattice configuration.

When a spin echo pulse is absent, the  dynamics of $\ell$ is dominated by single-particle processes and depends on the lattice configuration and atomic density distribution. For a lattice fully filled with atoms frozen at  discrete lattice sites, it can be described as:
\begin{eqnarray}
\frac{\ell(t)}{24}=\frac{\cos^5(B_Qt/\hbar)}{\gamma_x\gamma_y\gamma_z t^3}|\sin(\frac{\gamma_x t}{2})\sin(\frac{\gamma_y t}{2})\sin(\frac{\gamma_z t}{2})|
\label{subeq:lnoecho}
\end{eqnarray}
where $\gamma_{x,y,z}=b_{x,y,z}L_{x,y,z}/\hbar$. It is straightforward to estimate from Eq.~(\ref{subeq:lnoecho}) that under typical  experimental conditions, $\ell$ already decays significantly within $5$ms. In comparison, when a spin echo pulse is applied, up to second order in time we find
\begin{eqnarray}
\ell(t)&=&3-\frac{3t^2}{16\hbar^2}(40B_Q^2+27V_{\rm eff}^2)+O(t^3)
\label{subeq:lecho}
\end{eqnarray}
which shows that the effect of ${\bf b}$ is removed as a result of the echo pulse at second order in $t$. In this case the decay of $\ell$ mainly comes from  dipolar interactions and happens at a rate  much slower than the one without echo. It is worth to note that while for Ising interactions, such as those between dressed Rydberg atoms \cite{isingechokaden}, spin echo technique can completely cancel the effect of inhomogneous field ${\bf b}$, for dipolar interactions, there is a small residual effect due to the noncommutativity between different terms in Eq.~(\ref{subeq:H}), which can show up at longer times.

Since as indicated by Eq.~(\ref{subeq:lnoecho}) the dynamics of $\ell$ without a spin echo strongly depends on the lattice size, in the numerical calculation, we use a large lattice that is closer to real experimental conditions, where atoms populate a shell of outer radius $r_{\rm out}\approx 23d$ and inner radius $r_{\rm in}\approx 18 d$, with $d=266$nm the smallest lattice spacing \cite{Lepoutre2019}. At unit-filling, this volume includes  $N\approx 10^4$ atoms, which is much larger than the system size investigated before~\cite{Lepoutre2019}. Although the quantum many-body dynamics for such a large system can still be  simulated with our GDTWA approach, we find an alternative approach can significantly reduce the computational cost while capturing the collective spin dynamics in this case. In the alternative approach, we use a Gaussian sampling of three spin-variables $\sum_{\bf r}\langle \hat{S}^{x,y,z}_{\bf r}\rangle$ which neglects intra-spin correlations as
described in Ref.~\cite{Zhu_2019}. While this approach does not correctly account for the effect of $B_Q$, for this case without the echo pulse the effect of $B_Q$ is much weaker compared to the effect of inhomogeneities, and the time evolution of $\ell$ can nevertheless be well captured. The result is plotted in Fig.~3 by the blue dashed line, which agrees well with experimental data. We note that throughout this work, the gaussian TWA approach is only used for solving the time evolution of $\ell$ in the absence of an echo pulse, while all other simulations are performed with GDTWA.

\begin{figure*}
\centering
\includegraphics[width= 0.8\textwidth]{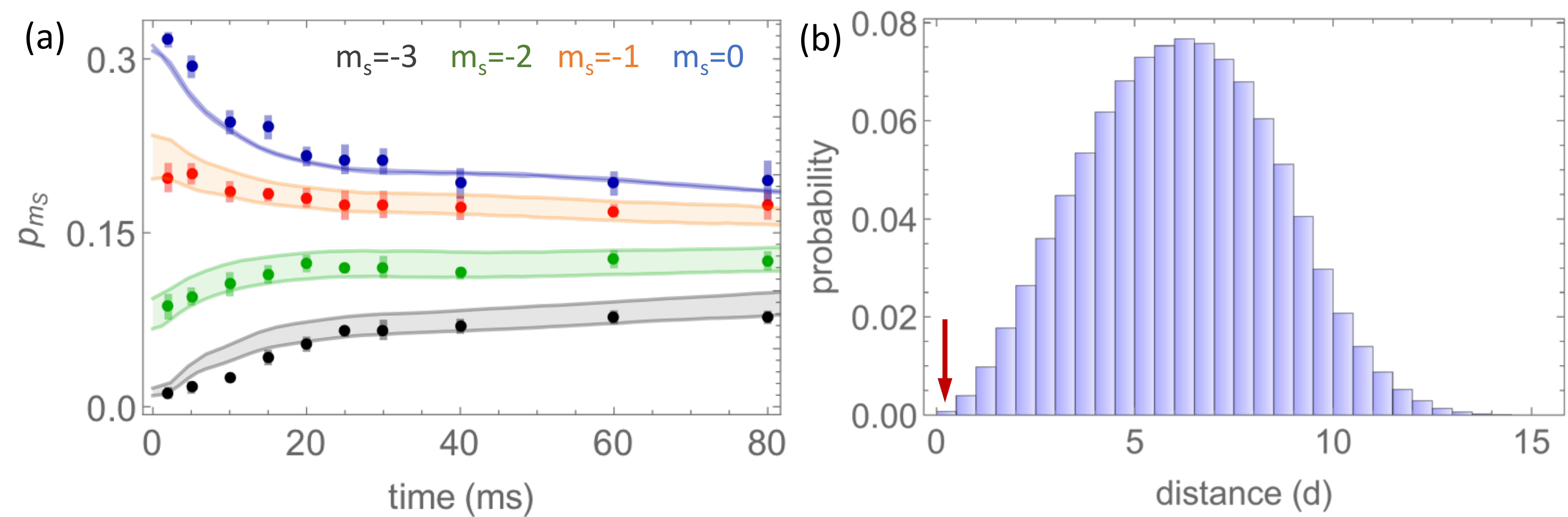}
\caption{\setlength{\baselineskip}{6pt} { (a) Dynamics of spin population calculated with GDTWA (lines) using the same model and parameters as for the blue solid line in Fig.~3. Different colors represent dynamics for different spin components: $m_s=-3$, $m_s=-2$, $m_s=-1$, $m_s=0$  from bottom to top. Color bands account for imperfection in the first pulse as in Fig.~1. (b) Distribution of inter-particle spacing in the GDTWA simulation, in the units of $d$. The leftmost bin (indicated by the red arrow) shows the fraction of atoms with distance closer than $r_{\rm cutoff}$, which is very small. Practically, in the GDTWA simulation all distances smaller than $r_{\rm cutoff}$ are set to be equal to $r_{\rm cutoff}$. This is partially motivated by the intuition that the contact interaction between Cr atoms under the experimental condition is very strong ($\sim$kHz \cite{Fersterer2019}) and thus prevents two atoms from being very close. Besides, since this fraction is tiny, except for small differences on the short-time scale ($\sim \hbar/18V_0\sim 5$ms, where $V_0=\frac{\mu_0(g\mu_B)^2}{4\pi d^3}$), the overall dynamics should remain similar for different choices of $r_{\rm cutoff}\le d$.  The atom number used in the simulation is $N=200$, which gives an overall density of $N/(L_xL_yL_z)\approx 0.6$.}}
\label{supfig:randpop}
\end{figure*}

 \begin{figure*}
\centering
\includegraphics[width= 0.8\textwidth]{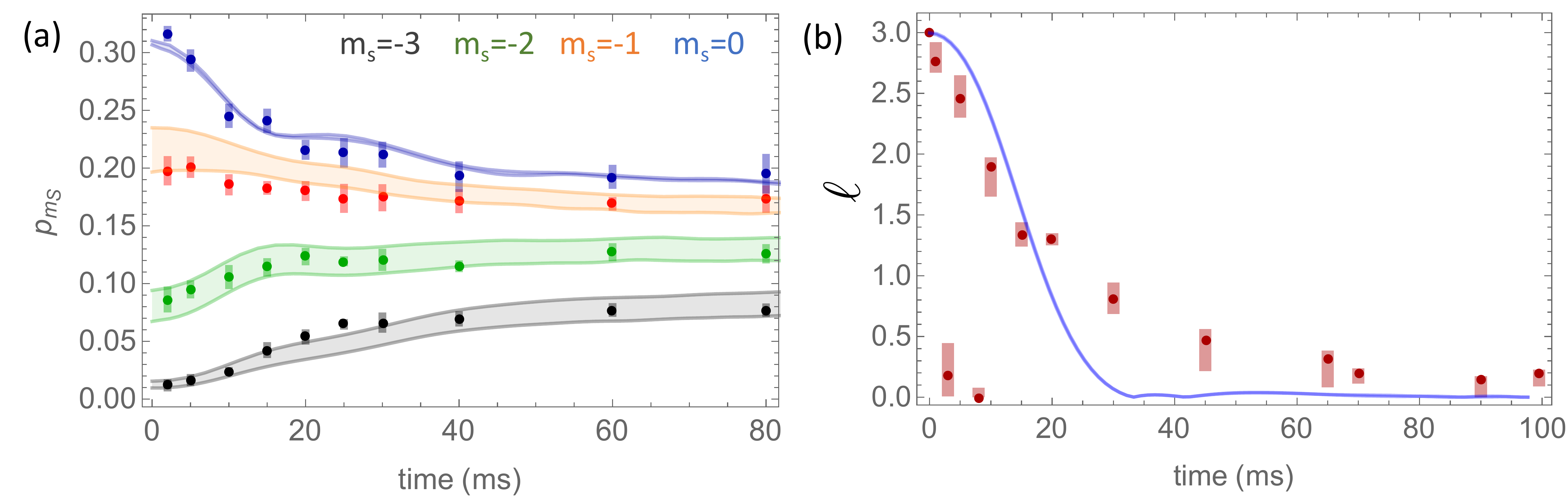}
\caption{\setlength{\baselineskip}{6pt} {Dynamics of spin population (a) and spin length (b) calculated with GDTWA (lines) for a regular lattice similar to the one in Fig.~1 (c) but with nonunity filling. To compare with the model accounting for atomic motion, the filling fraction is chosen to be $60\%$, close to the density used in Fig. \ref{supfig:randpop}. Lines in (a) plot dynamics for $m_s=-3$, $m_s=-2$, $m_s=-1$, $m_s=0$  from bottom to top. Color bands account for imperfection in the first pulse as in Fig.~1. A value $B_Q/h=-3$Hz is used for best describing the experimental spin dynamics $p_{m_s}$.}}
\label{supfigA}
\end{figure*}

For the case with an echo pulse applied, the spin dynamics is insensitive to the magnetic field gradients and thus does not strongly depend on the lattice size when the filling is unity. However, the theoretically calculated dynamics only captures the short-time dynamics well while shows significantly faster decay than experimental measurement at longer times [Fig.~3 black dashed line], where heating and tunneling effects becomes important. Although there is not an efficient way to exactly solve the quantum dynamics fully accounting for these effects and for sufficiently large system size, we investigate the spin dynamics including these effects in a phenomenological way. Instead of using a spin model where all atoms are frozen at discrete lattice sites, we allow a continuous random uniform distribution of ${\bf r}$ in a lattice of size $(L_x,L_y,L_z)$, while keeping atoms at least $r_{\rm cutoff}=0.5d$ away from each other [see Fig. \ref{supfig:randpop} caption]. Both the evolution of spin population [Fig\ref{supfig:randpop}(a)] and $\ell$ [solid line in Fig.~3] obtained with this model capture well the experimental observations.

 The  spin dynamics can  slow down  due to  non-unit filling of the lattice, which also introduces disorder in the spin  couplings, as was the case in Ref.~\cite{Yan2013}. In our experiment numerical simulations performed at lower filling fractions with frozen particles  indicate lower filling  is   not the cause of the observed slower contrast dynamics  in the experiment. To further illustrate this, in Fig. \ref{supfigA} we calculate the spin dynamics for a partially filled lattice, with atoms frozen at discrete sites. While the spin population evolves in a similar way as in Fig. \ref{supfig:randpop} and still describes experimental data, the calculated $\ell(t)$ significantly deviates from the experimental dynamics.

\begin{figure}
\centering
\includegraphics[width= 0.7\textwidth]{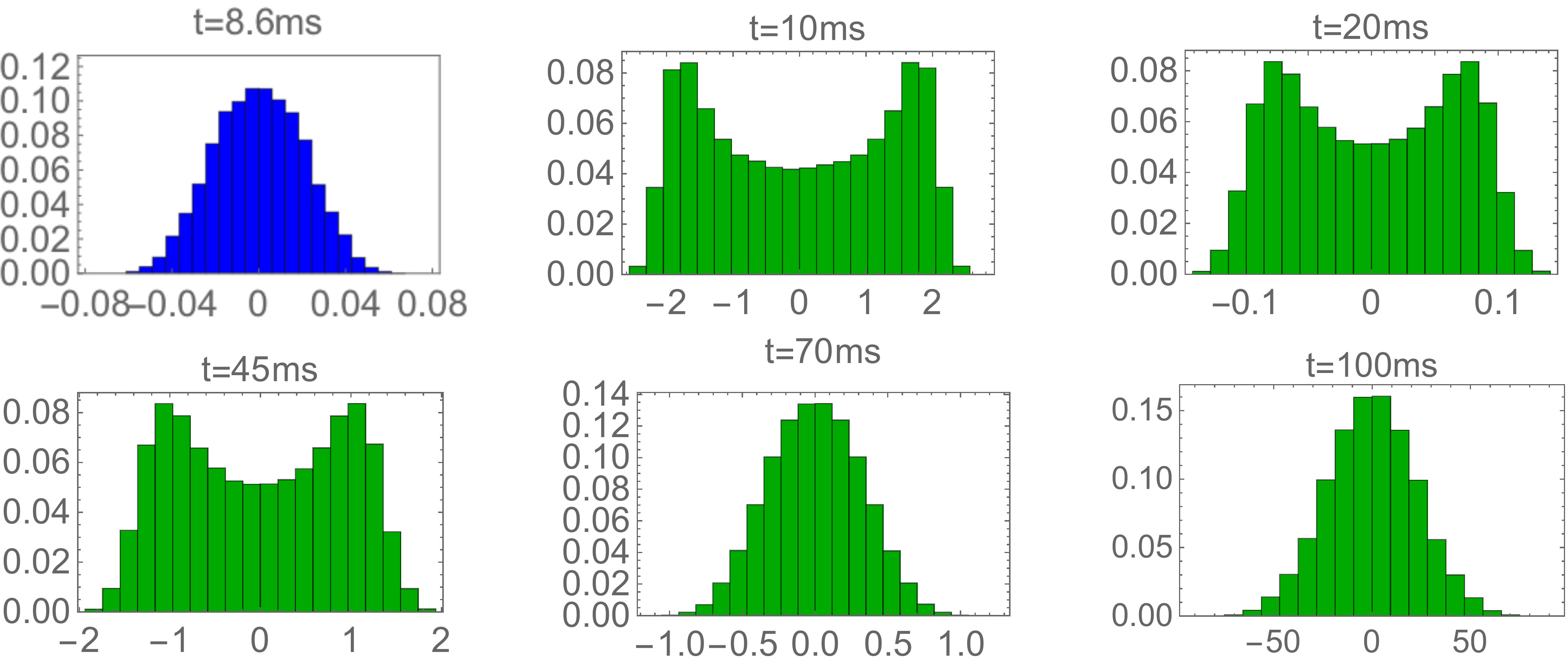}
\caption{\setlength{\baselineskip}{6pt} {Probability distributions of the normalized magnetization for the cases without (blue) and with (green) an  echo pulse, obtained with the model and parameters used in Fig.~3 for the blue dashed and solid lines, respectively. }}
\label{supfigB}
\end{figure}

In  Fig. \ref{supfigB}, we plot the numerically calculated probability distribution of the normalized magnetization for the toy model that accounts for heating and tunneling by allowing  particles  to be at continuous locations within the array. Our theoretical model also well describe the measured probability distributions (see FiG. 2 main text).

\begin{figure}
\centering
\includegraphics[width= 0.38\textwidth]{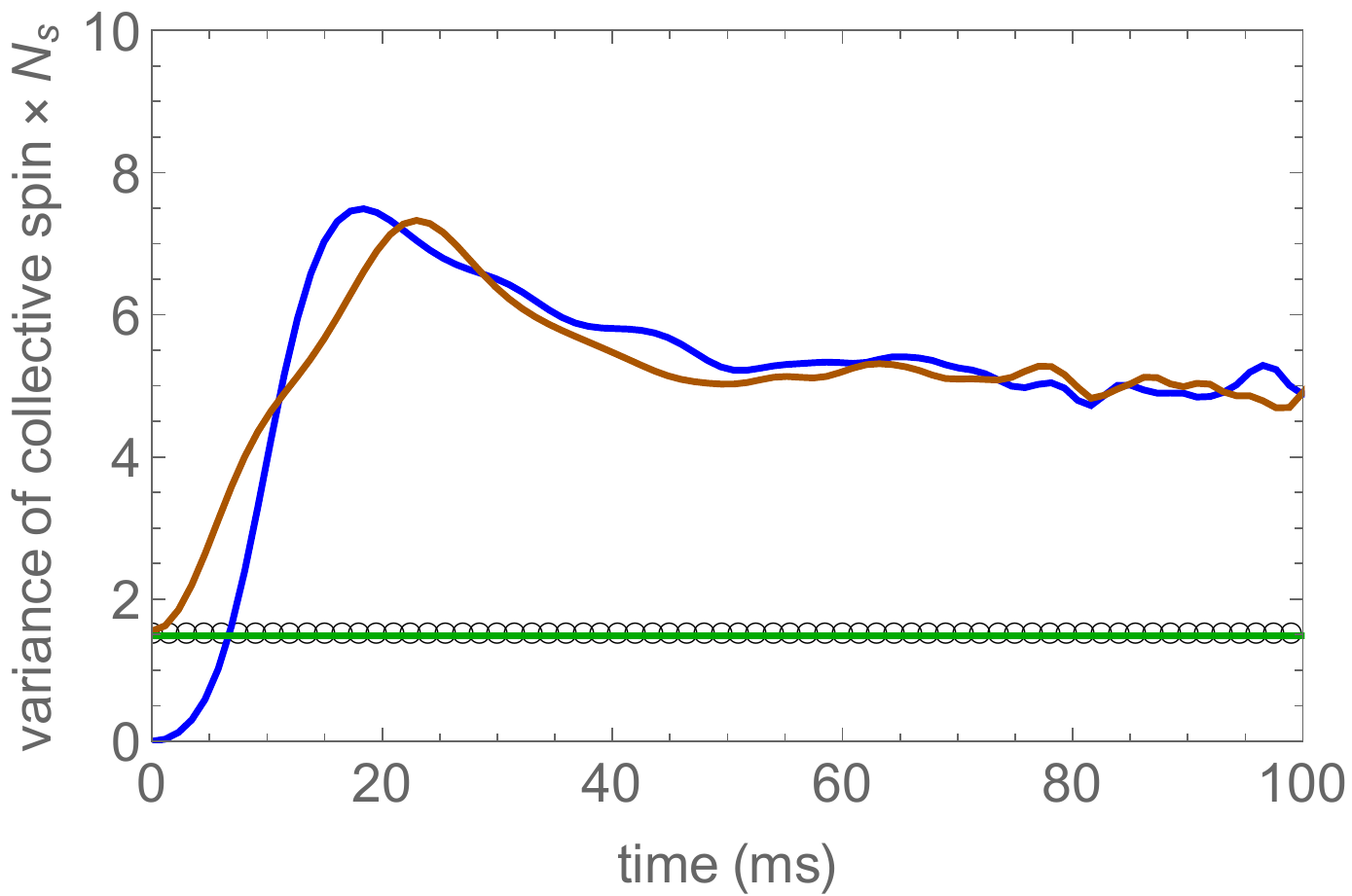}
\caption{\setlength{\baselineskip}{6pt} {Variance of collective spin projections, ${\rm Var({\hat j}_X)}$ (blue solid line),  ${\rm Var({\hat j}_Y)}$ (brown solid line), and ${\rm Var({\hat j}_z)}$ (green solid line), each multiplied by $N_s$. The results are obtained from GDTWA simulations for a unit-filled lattice with the same geometry as for Fig. 1(c), and with $B_Q/h=-5$Hz. The empty circles show the standard quantum noise  $3/2$ for $S=3$ atoms.  }}
\label{supfigVar}
\end{figure}

To further provide information on the quantum noise generated by strong interactions during the time evolution, in Fig.~\ref{supfigVar}, we use GDTWA to calculate the variance of collective spin projections, ${\rm Var({\hat j}_X)}$ for a unit-filled lattice, as used for Fig. 1(c).

\end{document}